\documentclass[twocolumn,nofootinbib,10pt]{revtex4-1}
\usepackage{graphicx,float}\usepackage[all]{xy}
\usepackage{amsmath,upgreek}
\usepackage{amssymb}
\usepackage{color}
\usepackage[colorlinks,hyperindex]{hyperref}

\newcommand{\bes}{\begin{subequations}}
\newcommand{\ees}{\end{subequations}}
\def\ben{\begin{eqnarray}}
\def\een{\end{eqnarray}}
\def\be{\begin{equation}}
\def\ee{\end{equation}}

\begin{document}

\title{Configurational entropy of  anti-de Sitter black holes }
 \author{Nelson R. F. Braga}
\affiliation{Instituto de F\'isica, Universidade Federal do Rio de Janeiro,
Caixa Postal 68528, RJ 21941-972 - Brazil}
\email{braga@if.ufrj.br}
\author{Rold\~ao da Rocha}
\affiliation{Centro de Matem\'atica, Computa\c c\~ao e Cogni\c c\~ao, Universidade Federal do ABC - UFABC\\ 09210-580, Santo Andr\'e, Brazil.}
\email{roldao.rocha@ufabc.edu.br}


\begin{abstract} 
Recent studies indicate that the configurational  entropy is an useful  tool to investigate the stability and (or) the relative dominance of states for diverse physical systems. Recent examples comprise the connection between the variation of this quantity and the relative fraction of light mesons and glueballs observed in hadronic processes.  
Here we develop a technique for defining a configurational entropy for an AdS-Schwarzschild black hole.  The achieved result  corroborates consistency with the Hawking-Page phase transition. Namely, the dominance of the  black hole configurational entropy will be shown to increase with the temperature.  In order to verify the consistency of the new procedure developed here,  we also consider the case of black holes in flat space-time. For such a black hole, it is known that evaporation leads to instability.
The configurational entropy obtained for the flat space case is thoroughly consistent with the physical expectation. In fact, we show that the smaller the black holes, the more unstable they are. So, the configurational entropy furnishes a reliable measure for  stability of black holes.

\end{abstract}


\maketitle

\section{ Introduction }

{A black hole is a particular solution of Einstein equations and 
has both a temperature related to its surface gravity and entropy, this last one associated with 
the black hole area. Black holes can radiate and evaporate, due to particle creation and annihilation, quantum fluctuations 
 near  the event horizon and also due to tunnelling methods across it. 
AdS-Schwarzschild black holes can be thermodynamically stable and were shown to 
manifest  a first-order phase transition, due to a negative free energy at high temperatures \cite{Hawking:1982dh}. 
The study of black holes in anti-de Sitter (AdS) space was, afterwards, refreshed in the AdS/CFT correspondence, 
comprising a relationship between string theory in a 5-dimensional AdS space and Yang-Mills theories on the AdS space
conformal boundary. Subsequently, the Hawking-Page phase transition was understood 
in the gauge theory setup as a  phase transition, from confinement to deconfinement \cite{Witten:1998qj,Witten:1998zw}.}

One of the most striking  applications of the informational entropy in physics is to 
provide a remarkable tool to analyze compact astrophysical  objects. In fact, a stability bound for stellar distributions was obtained in the context of the configurational entropy, both in  Refs. \cite{Gleiser:2013mga,Gleiser:2015rwa} and in Ref. \cite{Casadio:2016aum} as well. This last result provided a relevant procedure to scrutinize the stability of Bose-Einstein condensates of long-wavelength gravitons in a black hole quantum portrait, in the configurational entropy setup \cite{Casadio:2016aum}.  In both these developments, the critical stability region of stellar configurations have been consistently defined in complementary, distinct, paradigms that match observational data \cite{Gleiser:2013mga,Casadio:2016aum}. Hence, the  critical stellar densities were shown to match the Chandrasekhar stellar critical density, associated with critical points of the informational entropy.

\par
  The present day concept of  configurational  entropy has roots in early studies of informational entropy by  Shannon \cite{shannon},  introduced  in the context of  communication theory.  The earliest main idea was  to associate  different messages to distinct probabilities. When  $N$ messages are equally probable, Shannon defined the informational entropy of an unread message as being $\log_2 N = k_S \log N$, with  $k_S = 1/\ln 2$ playing a role analogous to the Boltzmann constant.
Entropy may be seen as a measure of the information deficit about a system. 
The informational entropy was re-introduced,  as the configurational entropy, in the recent literature by Gleiser and collaborators, in the context of complex systems and also compact astrophysical  objects, among other models \cite{Gleiser:2013mga,Gleiser:2015rwa,Gleiser:2011di,Gleiser:2012tu}. It represents a quantitative measure of an entropy of shape, based on successful scrutiny of some physical systems supporting local interactions.  The core of the configurational entropy is constituted by the modal fractions in informational entropy, emulating the collective coordinates-to-structure factor ratio in the thermodynamical entropy setup  \cite{Bernardini:2016hvx}. 

The configurational  entropy, also known as informational entropy\footnote{Ref. \cite{Bernardini:2016hvx} also discusses 
the configurational  entropy as a particular case of the conditional relative entropy \cite{Gleiser:2014ipa}, to better provide a formal 
study of the integral that defines the configurational  entropy.}, has been recently turned into a promising  tool for investigating the stability and/or the dominance of states in some physical models. 
For example, the configurational entropies of glueball and meson states, as described in the AdS/QCD framework, were  recently studied in \cite{Bernardini:2016hvx,Bernardini:2016qit}, providing prominent applications in the study of hadronic processes. A clear relation between the configurational entropy and the abundance  of the states was derived, wherein the states with smaller configurational entropy are more likely found in hadronic processes \cite{Bernardini:2016hvx,Bernardini:2016qit}.

The purpose of the present paper is to develop an approach for a consistent definition of the configurational entropy for an AdS-Schwarzschild  black hole.  We want to verify whether the Hawking-Page transition is appropriately represented in the configurational entropy setup. 
As a by-product, and as a check of consistency, we will also consider a black hole in flat space, for which the evaporation by  Hawking radiation leads to the black hole instability. This paper is organized as follows: Sect. II is committed to 
briefly revisit the configurational entropy setup and Sect. III is devoted to the construction of a configurational entropy for the AdS-Schwarzschild black hole. In Sect. IV the flat space case is accomplished and analyzed. Sect. V encloses the concluding remarks and outlook. 

\section{configurational entropy framework}

The configurational entropy is defined from the modal fraction, that is constructed upon of the Fourier 
 transform of the energy density associated with a  physical system \cite{Gleiser:2013mga}. 
 The starting point is the informational entropy, originally  defined for any system with  $N$ discrete modes by the expression $S_c = -{{\sum_{i=1}^N}}\;f_i\ln(f_i),$ where the $\{f_i\}$ are probability density functions \cite{shannon}.  For the case of a black hole, the energy density is assumed to be a function of the position $ \rho = \rho ({\vec r})$.  In what follows, all integrals are defined over the whole space.  In a space of $n$ spatial dimensions, the Fourier transform  of the energy density, 
\begin{equation}
{\tilde \rho} ({ \vec \omega})=(2\pi)^{-n/2}\!\! \int  \!\rho({\vec r}\,)\,e^{i\vec\omega \cdot{\vec r}} d^n  r\,\,,
\label{collectivecoordinates}
\end{equation} can be thought as the continuum limit of the collective coordinates in statistical mechanics, $\rho({\vec  r})=\sum_{j=1}^N \exp\left({-i{\vec \omega}_j \cdot {\vec r}}\,\right)\tilde\rho({\vec \omega}_j)$ \cite{Bernardini:2016hvx}.    The structure factor, $
s_N = \frac1N\sum_{j=1}^N\;\langle\;\vert\;\tilde\rho({\vec\omega}_j)\;\vert^2\rangle\,,$ normalizes the correlation of collective coordinates, as  
\begin{equation}
f({\vec \omega}_N)=\frac{1}{s_NN}{\langle\;\vert\;\tilde\rho({\vec\omega}_N)\;\vert^2\rangle}.\label{collective1}
\end{equation} 
The structure factor measures  fluctuations in the energy density, providing the system profile towards homogenization. The $N \to \infty$ limit yields  the modal fraction to read:
 \begin{equation}
 \label{modalfrac}
 f({\vec  \omega}) \equiv  \frac{\langle\;   \vert {\tilde \rho} ( {\vec \omega} ) \vert^{2} \rangle  }{ {\cal N} } \,,
 \end{equation} 
 
 \noindent where 
 \begin{equation}
 {\cal N}  \,=\,   \int \langle\;\;\vert \left({\tilde  \rho} (\vec\omega )\right)^* {\tilde  \rho} (\vec\omega )\vert\;\; \rangle d^n \omega \,.
 \label{norma}
 \end{equation}
 
 The configurational entropy is, then,  defined as  \cite{Gleiser:2011di,Gleiser:2012tu}
\begin{equation}
S_c[f] \,= \, - \int f(\vec\omega ) \ln f (\vec\omega )  d^n \omega 
\end{equation} 
 Critical points of the configurational entropy  imply that the system has informational entropy 
that is critical with respect to the {maximal modal fraction} $f_{\rm max}({\vec\omega})$, corresponding to 
 more dominant states \cite{Bernardini:2016hvx,Casadio:2016aum}.
The prototypical $d$-dimensional distribution is a Gaussian \cite{Gleiser:2011di}, wherein the configurational entropy estimates the information demanded in the reciprocal space to
 constitute the energy density in the position space. Such a distribution 
 represents an absolute minimum for spatially-localized
functions \cite{Gleiser:2011di}. Dynamical 5-dimensional kinks were also studied in Ref. \cite{Correa:2015vka}, in this context.

\section{AdS-Schwarzschild Black Holes and Information Entropy}
The thermodynamics of black holes in  AdS  space have been deeply studied in Ref. \cite{Hawking:1982dh}.
One of the relevant outcomes of this                                                                                                                                                                                                                                                                                                                                                                                                                                                                                                                                                                                                                                                                                                                                                                                                                                                                                                                                                                                                                                                                                                                                                                                                                                                                                                                                                                                                                                                                                                                                                                                                                                                                                                                                                                                                                                                                                                                                                                                                                                                                                                                                                                                                                                                                                                                                                                                                                                                                                                                                                                                                                                                                                                                                                                                                                                                                                                                                                                                                                                                                                                                                                                                                                                                                                                                                                                                                                                                                                                                                                                                                                                                                                                                                                                                                                                                                                                                                                                                                                                                                                                                                                                                                                                                                                                                                                                                                                                                                                                                                                                                                                                                                                                                                                                                                                                                                                                                                                                                                                                                                                                                                                                                                                                                                                                                                                                                                                                                                                                                                                                                                                                                                                                                                                                                                                                                                                                                                                                                                                                                                                                                                                                                                                                                                                                                                                                                                                                                                                                                                                                                                                                                                                                                                                                                                                                                                                                                                                                                                                                                                                                                                                                                                                                                                                                                                                                                                                                                                                                                                                                                                                                                                                                                                                                                                                                                                                                                                                                                                                                                                                                                                                                                                                                                                                                                                                                                                                                                                                                                                                                                                                                                                                                                                                                                                                                                                                                                                                                                                                                                                                                                                                                                                                                                                                                                                                                                                                                                                                                                                                                                                                                                                                                                                                                                                                                                                                                                                                                                                                                                                                                                                                                                                                                                                                                                                                                                                                                                                                                                                                                                                                                                                                                                                                                                                                                                                                                                                                                                                                                                                                                                                                                                                                                                                                                                                                                                                                                                 article was the discovery of a phase transition between the black hole 
space and the thermal AdS space without a black hole. For temperatures above a critical value, the black hole state is dominant, whereas  below the critical temperature the thermal AdS space state represents the dominat configuration. Although black holes are at thermal equilibrium with radiation,
they are not the preferred state below a certain critical temperature \cite{Hawking:1982dh}.
Subsequently to this investigation, but shortly after the discovery of the AdS/CFT correspondence   
\cite{Maldacena:1997re,Gubser:1998bc,Witten:1998qj}, the connection between the thermal  properties of an AdS-Schwarzschild  black hole and 
 those of a dual gauge theory were unveiled in  Ref. \cite{Witten:1998zw}.  The thermal phase transition found in Ref. \cite{Hawking:1982dh},  presently known as Hawking-Page transition,  was interpreted in Ref. \cite{Witten:1998zw} as a transition between a deconfined phase (corresponding to a high temperature) and a confined phase (low temperature). 
In order to understand the phase transition, one has to  compare the actions of the AdS metric and the AdS-Schwarzschild metric.  The phase transition occurs in the regime where the two actions 
are  equal  \cite{Hawking:1982dh}.  
 
 The anti-de Sitter space is a solution of  the vacuum
Einstein equation with a negative cosmological constant.  The metric of a (thermal) ($n+1$)-dimensional Euclidean AdS space,
 in global coordinates,  can be written as: 
\begin{equation}
\label{AdS}
 ds^2 \,\,= \,\, \left(1+\frac{r^2}{b^2}\right) \,d{\bar t}^{\, 2 }+ \frac{dr^2}{1+\frac{r^2}{b^2}} + r^2 d\Omega^2_{(n-1)} \,\,.
 \end{equation}
 The boundary of the space, at $r \to \infty$, is the product of  a $(n-1)$-sphere ${\boldmath S}^{n-1}$  and the circumference  $S^1$, associated with 
 the temporal variable that has a period $\bar \beta$, where the equivalence $\bar t \sim \bar t +\bar\beta$ defines the time circle $S^1$. 
 
 On the other hand, an AdS-Schwarzschild black hole metric, with the same asymptotic (large $r$) boundary reads
 \begin{equation}
 \label{AdSBH}
 ds^2 \,\,= \,\, \left(1+\frac{r^2}{b^2} \,- \frac{w_nM}{r^{n-2}}\right) \,dt^2 + \frac{dr^2}{1+\frac{r^2}{b^2}- \frac{w_nM}{r^{n-2}}} + r^2 d\Omega^2 \,\,,
 \end{equation}
 according to the notation in Ref. \cite{Witten:1998zw}, where the constant $w_n$ is related to the black hole mass $M$ and to the $n+1$-dimensional Newton constant $G_N $
 \begin{equation}
 \label{wn}
 w_n = \frac{ 16 \pi G_N}{(n-1) {\rm Vol} (S^{n-1})} \,,
 \end{equation} 
 where ${\rm Vol} (S^{n-1}) = 2\pi^{n/2}/\Gamma(n/2) \,$ is the volume of ${\boldmath S}^{n-1}$. 
 
  The location of the black hole horizon is the radial position where the time component of the metric vanishes and the radial component of the metric  becomes singular. Hence, the black hole outer region corresponds to $ r > r_h$, with $ r_h$  calculated as the largest root of the algebraic equation
  \begin{equation}
 \label{horizon}
1+ \frac{r^2}{b^2}  \,- \frac{w_nM}{r^{n-2}} \,= 0 \,.
  \end{equation}

 The configurational entropy of a system essentially depends  on the spatial distribution of the energy. The Fourier transform of the energy density is used to define the modal fraction in Eq. (\ref{modalfrac}).  
Therefore, our proposal is to associate some energy density with the black hole geometry and look for the corresponding configurational entropy. The energy density of such a  static configuration is related to the mass distribution. Hence, the first question to be asked is: does it make sense to talk about a mass distribution for the space outside a black hole?
 One  associates a black hole as the  geometry produced by mass that is inside the event horizon. Then, we argue argue whether  it would be meaningful to look for a quantity that represents a mass distribution, at least as a tool to find the configurational entropy,  just using the space outer of  the event horizon.  We claim that this does make sense, accordingly. We will base our argument on the fact that one can calculate the mass of the black hole from the action integral of the geometry outer to the event horizon:
  \begin{equation}
 \label{TotalMass}
  M = \frac{\partial I}{\partial \beta} \,,
  \end{equation}
 where the time variable is again periodic, of period $\beta$.  The action $I$ denotes a regularized form of the action integral associated with the black hole, to be subsequently defined, and $\beta$ is the period of the time variable $t$, that must be related to the position of the horizon by the condition \cite{Witten:1998qj} 
   \begin{equation}
 \label{tempXhorizon}
\beta =  \frac{4 \pi b^2 r_h }{nr_h^2 + (n-2) b^2}   \,,
  \end{equation}
 otherwise the geometry of Eq. (\ref{AdSBH}) would have a conical singularity at $ r = r_h$. The black hole temperature reads $T = 1/ \beta.$ 
 
Now we look for a local form for Eq. (\ref{TotalMass}) in order to identify a local mass density.
 The regularized action $I$ comes from the difference between the black hole and the thermal AdS actions. The Einstein gravity actions
 corresponding, respectively, to the geometries (\ref{AdS}) and (\ref{AdSBH})  are:  
 \begin{eqnarray} 
  I_{\rm AdS} &=&   \frac{n}{8\pi G_N}  \frac{ {\rm Vol} ( S^{n-1} )}{ b^2} \, \int_0^{\bar \beta} d\bar{t}  \int_{0}^R r^{n-1} dr \,\,,
  \label{THAdSAction}\\I_{\rm BH} &=&  \frac{n}{8\pi G_N}   \frac{ {\rm Vol} ( S^{n-1} )}{ b^2} \, \int_0^{\beta} dt \int_{r_h}^R r^{n-1} dr  \,\,,
  \label{BHaction}  
    \end{eqnarray}
 where $R$ is a large (regulator) radius
 that cancels out, when one calculates the difference between the above actions. 
 
 In order that  the actions have the same asymptotic geometry at $r\to \infty$, one must impose the condition that the sizes of the temporal circles equal each other:
\begin{equation}
 \label{relationoftimes}
\sqrt{ \left(1+ \frac{R^2}{b^2} - \frac{w_nM}{R^{n-2}} \right) }\,\,\,dt \,=\,  \sqrt{ \left(1+ \frac{R^2}{b^2}\right) } \,\,\,d\bar{t} \,\,.
 \end{equation}
 Then, the periods of the time coordinates are related by:  
 \begin{equation}
 \label{relationoperiods}
 {\bar \beta} \,=\, \beta \, \sqrt{ 1 \,-\, \frac{ w_nM}{ R^{n-2}\left(1+ \frac{R^2}{b^2}\right) }}
  \,\,.
 \end{equation}
 
 The regularized action $I$, that represents the effect of the presence of the black hole with respect to the AdS space without a black hole, reads:
\begin{eqnarray} 
I &=& I_{\rm BH} \,-\,  I_{\rm AdS} \nonumber\\&=& \frac{n}{8 \pi G_N}   \,\frac{ {\rm Vol} ( S^{n-1} )}{ b^2}\left( \left[ \beta \frac{r^n}{n}\right]_{r_h}^{R} - \left[ \beta' \frac{r^n}{n}\right]_{0}^{R} \right) \cr
&=& \frac{ {\rm Vol} ( S^{n-1} )}{ 4 G_N b^2} \, \frac{r_h^{n-1}(b^2\!- r_h^2) }{\left[ nr_h^2 \!+\! (n\!-\!2) b^2 \right] }.
  \label{Regularized}
 \end{eqnarray}    
 The difference of the actions changes sign at $r_h = b$.   For $ r_h > b$, the black hole action is smaller than the thermal AdS action. Consequently, the action $I$ is negative and the black hole is the most probable configuration. When $r_h < b$, the thermal AdS is the dominant.  This corresponds to the so called Hawking-Page phase transition, based upon a semi-classical approach.  
 
 The mass of the black hole is obtained in a straightforward way  from the  regularized action (\ref{Regularized}), by   using  Eq.  (\ref{TotalMass})  and  the expression:
 \begin{equation}
 \label{betaerrezero}
  \frac{\partial r_h}{\partial \beta } \,=\, \frac{1}{4 \pi b^2}  \frac{\left( nr_h^2 + (n-2) b^2 \right)^ 2}{  (n-2) b^2- nr_h^2} \,.
 \end{equation}
 
 Our goal  is to find an expression for the mass in terms of a mass density $\rho$:
\begin{equation}
 \label{ massdensity01}
M = \int_{0}^R r^{n-1} \rho (r)  dr   \,, 
\end{equation}
and then use the density  to find the configurational entropy.  There is a non-trivial point in the calculation presented in Eq. (\ref{Regularized}), that is of prominent  importance for our purposes. 
 Namely, the horizon position depends on the temporal period $\beta$, or equivalently, on the temperature. Hence, one can not commute the order of the differentiation with respect to $ \beta$, within the spatial integration, when calculating the mass using Eq. (\ref{TotalMass}) and the action integrals   of  (\ref{BHaction})  and (\ref{THAdSAction}), as well. 
 
 Hereon we will consider the case of a 4-dimensional AdS-Schwarzschild black hole, as studied in Ref. \cite{Hawking:1982dh}, by taking $n= 3$.  The  Fourier transform of the density  will then be used to find the modal fraction and, subsequently, the configurational entropy.  For a 3-dimensional function that depends only upon the radial coordinate, the Fourier transform takes the form:
\begin{equation}
 \label{ Furiermassdensity01}
{\tilde \rho}(\omega) \, =\, \frac{\sqrt{2}}{\omega \sqrt{\pi} } \lim_{R\to \infty }  \int_{0}^R r \rho (r)  \sin (\omega  r) dr   \,.
\end{equation}

 In order to find the density to be used, let us look first at the contribution of 4-dimensional black hole action to the mass:
 \begin{eqnarray} 
 \frac{ \partial  I_{\rm BH} }{\partial \beta } &=& \,\frac{3}{ 2 G_N} \frac{ \partial }{\partial \beta } \left[ \beta  \int_{r_h}^R r^{2} dr \right]  \,\,\nonumber\\&=& \,  \frac{3}{ 2 G_N}  \, \left[ \int_{r_h}^R r^{2} dr  \,+ \beta \frac{ \partial }{\partial \beta } \int_{r_h}^R r^{2} dr  \right]\,.
  \label{derivativebeta}
 \end{eqnarray}
  The second term in the brackets  can be calculated using the relation (\ref{tempXhorizon}), that relates the horizon position and $\beta$, yielding
  \begin{equation} 
 \beta \frac{ \partial }{\partial \beta } \left[  \int_{r_h}^R r^{2} dr \right]  \, = \,  \frac{ r_h^3 \left( 3 r_h^2 +  b^2 \right) }{ \left(  3 r_h^2 -  b^2 \right) } \,\,\equiv \,\,H( r_h) 
  \label{resultderivativebeta}
 \end{equation}
 
 The other relevant term that must be analyzed is one appearing in the contribution from the thermal AdS to the mass:
 \begin{equation} 
\!\!\frac{ \partial  {\bar \beta}  }{\partial \beta } = 1 -  \frac{ r_h^3 +  b^2  r_h   }{  2 R^3} 
  - \frac{ r_h}{2 R^3 } \frac{ \left( 3 r_h^2 + b^2 \right)^2 }{ \left(  b^2- 3 r_h^2 \right) } \,\equiv
 1 - \frac{ F( r_h) }{R^3 }
  \label{resultbetabetaprime}
 \end{equation}
 We can write the density in terms of a sum of three terms: 
  \begin{equation}
   \rho (r)  \,=\, \rho_{_{\rm Vol}}  +\rho_{_{\rm hor}} +  \rho_{\infty} \,,
\end{equation}
where the volumetric term $\rho_{_{\rm Vol}} $  corresponds to the sum of  the first part of Eq. (\ref{derivativebeta}) and the unit in Eq. (\ref{resultbetabetaprime}), multiplied by the spatial part of the thermal AdS action, namely, 
\begin{equation}
\rho_{_{\rm Vol}} \,=\,- \, \frac{3}{2  G_N}    \,\,,   \,\,\, \textrm{ for }\,\,   ( 0 \le  r  \le r_h \,)\,\,.
\label{volumedensity}
\end{equation}
The corresponding Fourier transform reads:
\begin{equation}
{ \tilde {\rho}}_{_{\rm Vol}} \,=\,- \frac{3}{ \sqrt{2\pi}\omega^2 G_N} \left[ \frac{\sin (\omega r_h ) }{\omega} - { r_h \cos (\omega r_h ) } \right] \,.
 \end{equation}

 The term $\rho_{\infty}$ corresponds to the other contributions of Eq. (\ref{resultbetabetaprime}) to the thermal AdS action. It can be associated with a density   
  \begin{equation}
\rho_{_\infty} \,=\,  \frac{ 3F( r_h) }{2 G_N R^3}   \,.
\end{equation}
The Fourier transform for  this term is 
\begin{equation}
{ \tilde {\rho}}_\infty \,=\frac{3F( r_h)}{ \sqrt{2\pi} G_NR^3}  
 \left[ \frac{\sin (\omega R) }{\omega} - {R \cos (\omega R) }\right]\,,
 \end{equation}
that vanishes when one takes the limit $R \to \infty$. 
The remaining part of the density $ \rho_{_{hor}} $ is the one that corresponds to the horizon term in Eq. (\ref{resultderivativebeta}). Since this term is associated with the dependence of the horizon position on the temperature, one could think that a possible representation could be density localized at this position, like
\begin{equation}
 \rho_d \,=  \, \frac{3 }{2 G_N}   \frac{ H( r_h) }{r^2}  \, \delta ( r - r_h ) \,.   
\end{equation}
 However such a term  would lead to a Fourier transform that is not square integrable. Moreover, it does not reflect the physical situation where the horizon is a coordinate singularity, rather than an essential singularity. Therefore, the energy density should not be singular, at this location.  It turns out that a density that is consistent with 
this physical situation, in the sense of leading to a square integrable Fourier transform and to a finite configurational entropy, reads:
\begin{equation}
\rho_{_{\rm hor}}\,= \, \frac{3   H( r_h)}{ G_N  \, r_h^2  }   \, \,  \frac{ 1 }{ \,r}   \,\,\, \textrm{ for }\,\,   ( 0 \le  r  \le r_h \,)\,\,,
\end{equation}
with corresponding Fourier transform
\begin{equation}
{ \tilde {\rho}}_{_{\rm hor}} \,=  \frac{6}{ \sqrt{2\pi} \,  G_N  }  \, \frac{  H( r_h) }{r_h^2 }  
 \left[ \frac{ 1 -  \cos (\omega r_h  ) }{\omega^2} \right]\,.
 \end{equation}
 
It is interesting to remark that although the term $\rho_{_{\rm Vol}}$ in eq. (\ref{volumedensity}) is negative, it is straightforward to show that the sum of all terms  $\rho_{_{\rm Vol}} + \rho_{_\infty} + \rho_{_{\rm hor}} $ is positive everywhere.
 
  Now that we have the Fourier transforms of the energy densities, the  modal fraction can be derived, as defined in Eq. (\ref{modalfrac}):  
 \begin{equation}
 f(\omega) \,=\, \frac{1}{\cal N}  \vert   \tilde{\rho}_{_{\rm Vol}}  +\tilde{\rho}_{_{\rm hor}}  \vert^2, 
 \end{equation}
 where $\cal N$ is the normalization factor for this 3-dimensional case:
 \begin{equation}
 {\cal N }\,=\, 4 \pi   \int_0^\infty \,  \vert  {\tilde \rho}_{_{\rm Vol}}  + { \tilde\rho}_{_{\rm hor}}   \vert^2 \omega^2 d\omega\,.
 \end{equation}
 The configurational entropy is then 
\begin{equation}
\label{CEAdS}
 S_c(r_h)\,=\, -   4 \pi \int_0^\infty \,  \ln \vert  f(\omega) \vert \,  f(\omega)  \omega^2 d\omega\,.
 \end{equation}
The minimum value of the temperature for the 4-dimensional AdS black hole solution considered here 
corresponds to $r_h = b/\sqrt{3}$, that is the smallest possible value for the horizon radius, 
 for the large black hole solution considered here. We  numerically calculated the configurational entropy given in Eq. (\ref{CEAdS}) for the  illustrative range $ 0.6 \le  r_h/b  \le  6$.
 The results  are shown in  Fig. 1. We found a monotonic decrement of the configurational entropy with the event horizon radius.
 
 \begin{figure}[H]
\centering\includegraphics[width=7.9cm]{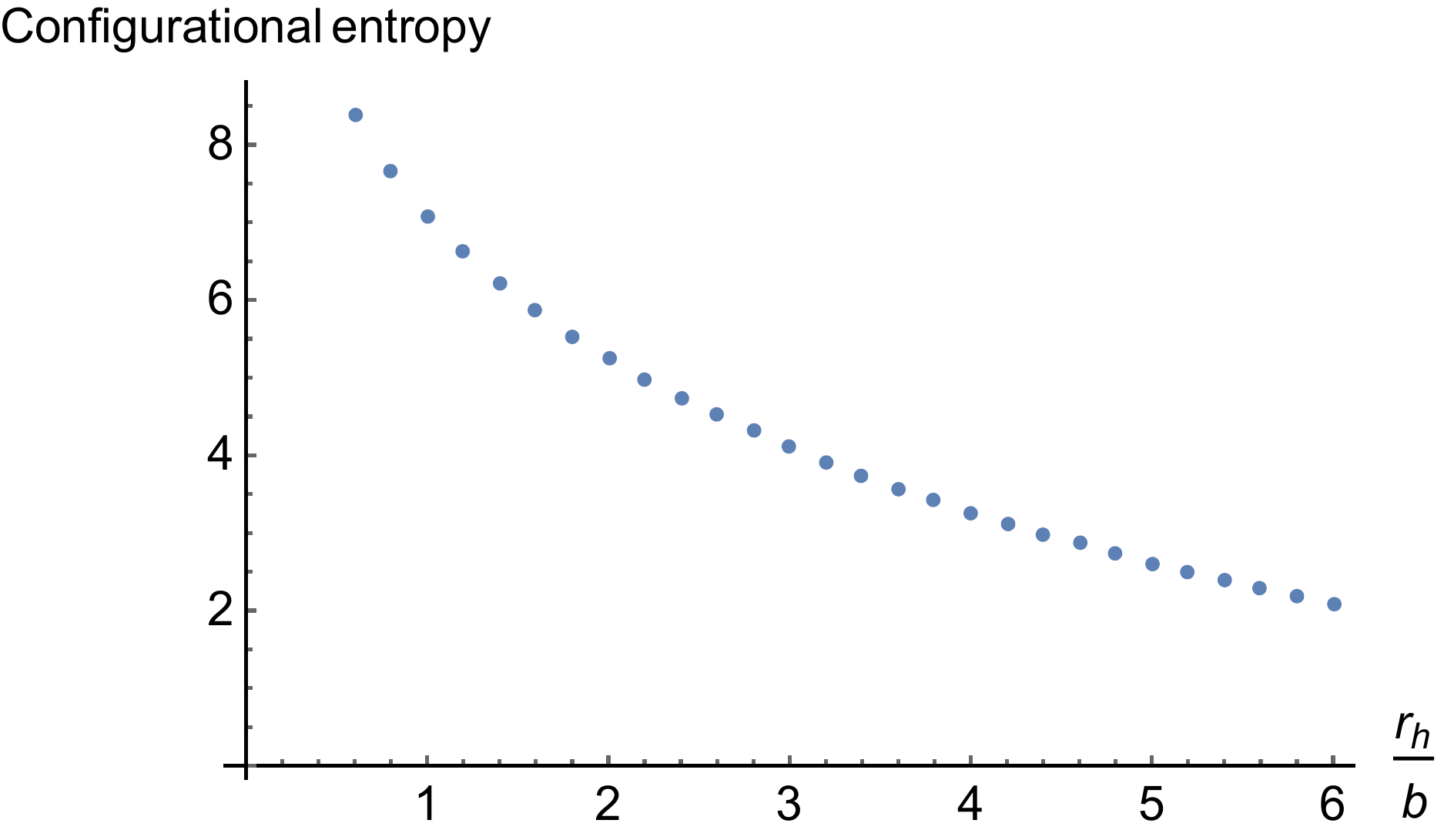}
\caption{Configurational entropy for the AdS-Schwarzschild black hole as a function of the event horizon in AdS radius units.}
\label{f1}
\end{figure}
 
 It is expected that,  due to the Hawking-Page transition, the black hole states are more  dominant, for temperatures above the critical one, corresponding to $ r _h/b = 1$.  Moreover, the dominance is related to the difference between the black hole action and the thermal AdS action. The semi-classical argument, underlying the Hawking-Page transition, consists of the relative probability of a given configuration that is proportional to the exponential of (minus) the corresponding action. Hence, for the 
 black hole geometry,  the probability of finding this space is governed by the factor $\exp \left(\, - I  \, \right)$, where $I$ is the difference between the black hole action and the thermal action, computed in Eq. (\ref{Regularized}).
  The larger the ratio $r_h/b$,  the bigger the difference between the actions is. 
 Consequently,  the higher the values of  $r _h/b$, the more dominant  the black hole state becomes. 
 Thus, the  obtained result for  the configurational entropy is totally consistent to the semi-classical picture.
  The entropy decreases as the ratio  $ r _h/b >  1$ increases, indicating more dominance.  Then, from the point of view of the configurational entropy, we found a result that shows that, in the region of horizon radius $ b/\sqrt{3} <  r_h  $, the black hole space becomes more stable as the radius increases. There is no particular signature of the  Hawking-Page transition point from the point of view of the configurational entropy.  
  
By taking  the logarithm of the configurational entropy as a function of the black hole radius, the possibility of a scaling relation between the configurational entropy and the AdS-Schwarzschild black hole event horizon can be inferred, illustrated in Fig. 2 below.  
 \begin{figure}[H]
\centering\includegraphics[width=7.9cm]{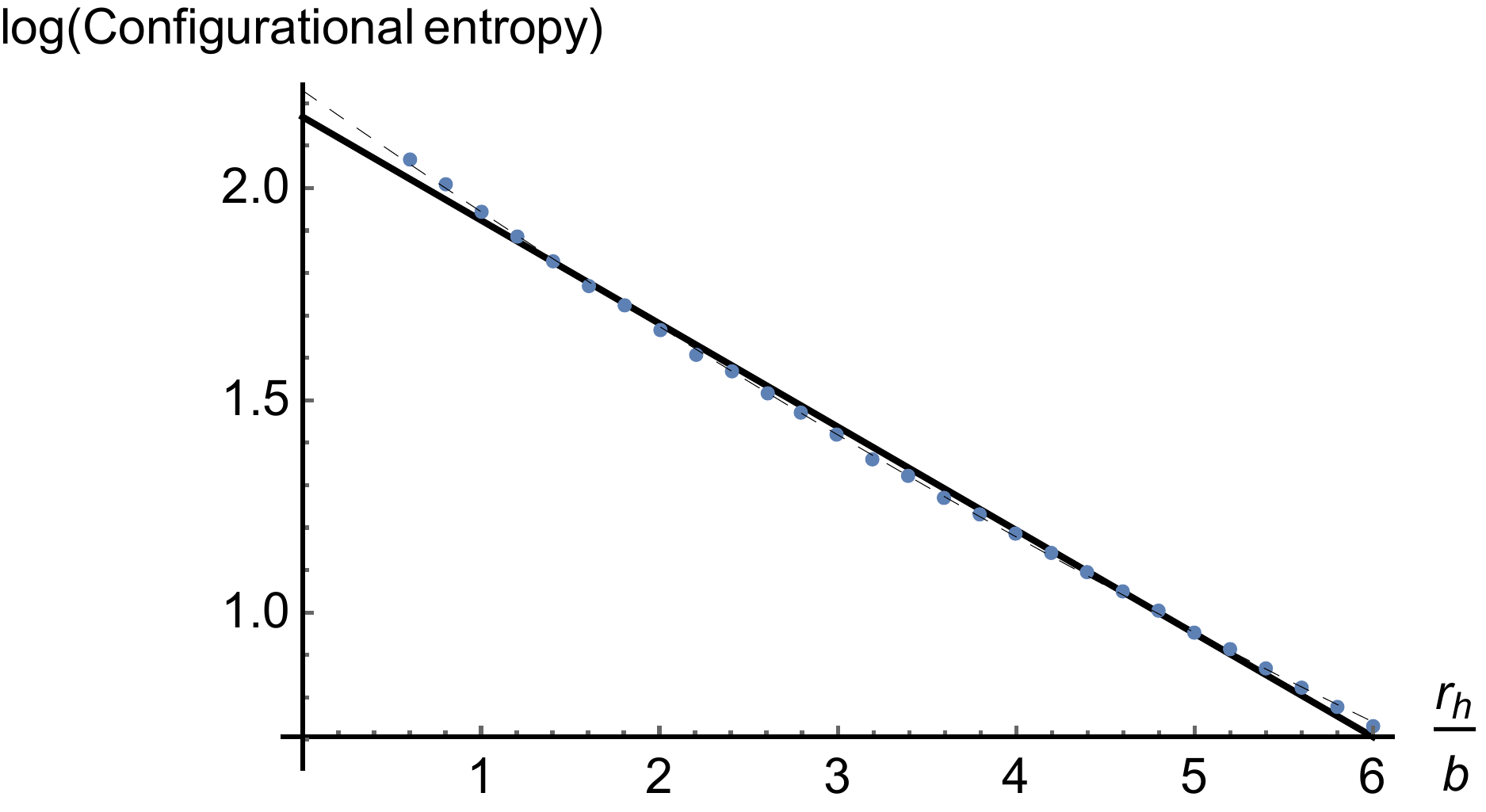}
\caption{Log of the configurational entropy for the AdS-Schwarzschild black hole as a function of the event horizon in AdS radius units (gray list plot); the black line refers to the linear regression and the dashed line represents a quadratic fit.}
\label{f55}
\end{figure}
 The linear regression makes the logarithm of the configurational entropy at the list plot in Fig. 2 to be expressed 
as the monic polynomial  $2.1676 - 0.2434 \frac{r_h}{b}$. Moreover, the list plot can be adjusted by the quadratic fit $2.2298 - 0.2929 \frac{r_h}{b} + 0.0065 \frac{r_h^2}{b^2}$. Due to the tiny value of the quadratic coefficient, the linear regression 
is a good attempt to a scaling relation between the configurational entropy and the AdS-Schwarzschild black hole event horizon radius, with an error of 6.1\%,  in the range $1\lesssim \frac{r_h}{b}\lesssim 6$.
  
\section{ Flat space case}
As a test of  consistency for the approach developed in the previous section for calculating the configurational entropy,  let us see what happens if ones applies a similar analysis for the flat space case.
We know that Schwarzschild  black holes in flat space evaporate by emitting Hawking radiation. The smaller the black hole, the shorter its life time is. Hence, the stability of such a geometry increases with the mass.

A 4-dimensional (Euclidean) Schwarzschild black hole in flat space can  be described by 
\begin{equation}
\label{SCHBH}
 ds^2 \,\,= \,\,  S(r)  \,dt^2 + \, \frac{  dr^2 } {  S(r) }\,  + r^2 d\Omega^2 \,\,.
 \end{equation}
 where the Schwarzschild factor reads \begin{eqnarray} S(r) \equiv 1 - \frac{2G_NM }{r} \,.\end{eqnarray}

 The horizon is located at $r = r _h = 2 G_N M$ and the temperature is the inverse of the period $ \beta $ of the temporal coordinate, related to the mass $M$  by: 
 \begin{equation}
 T =  \frac{ 1 }{ 8 \pi G_N M } \,.
 \end{equation}
 
The approach developed  in the previous section for AdS space consisted of starting with the action for the outer space to the black hole horizon and then expressing the mass of the black hole with respect to the volume integral of a ``mass density''.
The used mass density does not represent the actual mass distribution, concerning the inner region to the horizon. It is just a model for finding a consistent definition for the configurational entropy. 

For the case of the metric in Eq. (\ref{SCHBH}), if we try to find the mass of the black hole from the Einstein-Hilbert action,  corresponding to the outer region to the horizon, we face the following situation: the Ricci curvature scalar equals zero in this region.
Therefore, there is no volume term in the action integral.  However, for a region with a boundary, the gravity action has a  surface contribution.  The so called Gibbons-Hawking term \cite{Gibbons:1976ue} (see also \cite{BallonBayona:2007vp}  for the AdS  space case).  Using usual spherical coordinates, with $ d\Omega^2 = \sin \theta d\theta d \phi$, it takes the form
\begin{equation}
\label{GH} 
I_{\rm GW} \,=\, -\,  \frac{1}{ 4 \pi G_N}  \int_0^{\beta} dt \int_0^{2\pi} \!\!\!\!\int_0^\pi  \sqrt{h} K \,d\theta \, d\phi 
 \end{equation}
 where the integral is performed at the boundary (the event horizon, in this case), $h$ is the induced metric at $r = r_h $  and $K$ is the trace of the extrinsic curvature of space. The form of $K$ is 
  \begin{equation}
\label{K}  
K \,=\, \frac{1}{\sqrt{g} }  \, \partial_m \left( \sqrt{g} {\hat u}^m \right) \,,
 \end{equation}
where $ {\hat u}^m$ is the unit vector normal to the horizon that has only a radial component  $ {\hat u}^r = - \sqrt{ S(r) } $,   such that $g_{rr} {\hat u}^r {\hat u}^ r \,=1 $.  It yields 
 \begin{eqnarray}
\label{Kresultado}  
\left[ \sqrt{h} K \right]_{\rm horizon}\,&=&\, \left[\sqrt{S(r) } \, r^2 \sin\theta \, \frac{1}{\sqrt{g} }  \, \partial_m \left( \sqrt{g} {\hat u}^m \right) \large\right]_{\rm horizon}\nonumber\\&=& - M G_N \sin\theta  \,,
 \end{eqnarray}
implying that
\begin{equation}
\label{finalGH} 
I_{\rm GW} \,=\, \,  \beta \, M  \,.
 \end{equation}

Therefore, Eq.  (\ref{TotalMass}) holds also in the asymptotically flat space case. The mass 
in this case comes from the derivative of the Gibbons-Hawking boundary term with respect to  $\beta $. 
This surface term plays a role similar to the horizon contribution of the AdS volumetric action. In the AdS case, there is 
 a contribution to the energy that comes from the derivative of the radial limit of the integration (the horizon position) with respect to $ \beta$. Proceeding in an analogous way, we can represent this energy that is localized at the horizon,  by a density of energy that is square integrable and singular at the origin: 
\begin{equation}
\rho_{_{\rm Flat}}\,= \,  \frac{2 M }{ r_h^2 \, r   }  \,=\, \frac{ 1}{ r \, r_h \,G_N}  \,\,\, \textrm{ for }\,\,   ( 0 \le  r  \le r_h \,)\,\,.
\end{equation}
The corresponding Fourier transform yields
\begin{equation}
{ \tilde {\rho}}_{_{\rm Flat}} \,=\,  \frac{ \sqrt{2} }{ \sqrt{\pi } \,  r_h \, G_N  }
 \left[ \frac{ 1 -  \cos (\omega r_h  ) }{\omega^2} \right] \,.
 \end{equation}

The results for the configurational entropy of the flat space black hole are shown in Fig. 2. 

 \begin{figure}[H]
\centering\includegraphics[width=7.9cm]{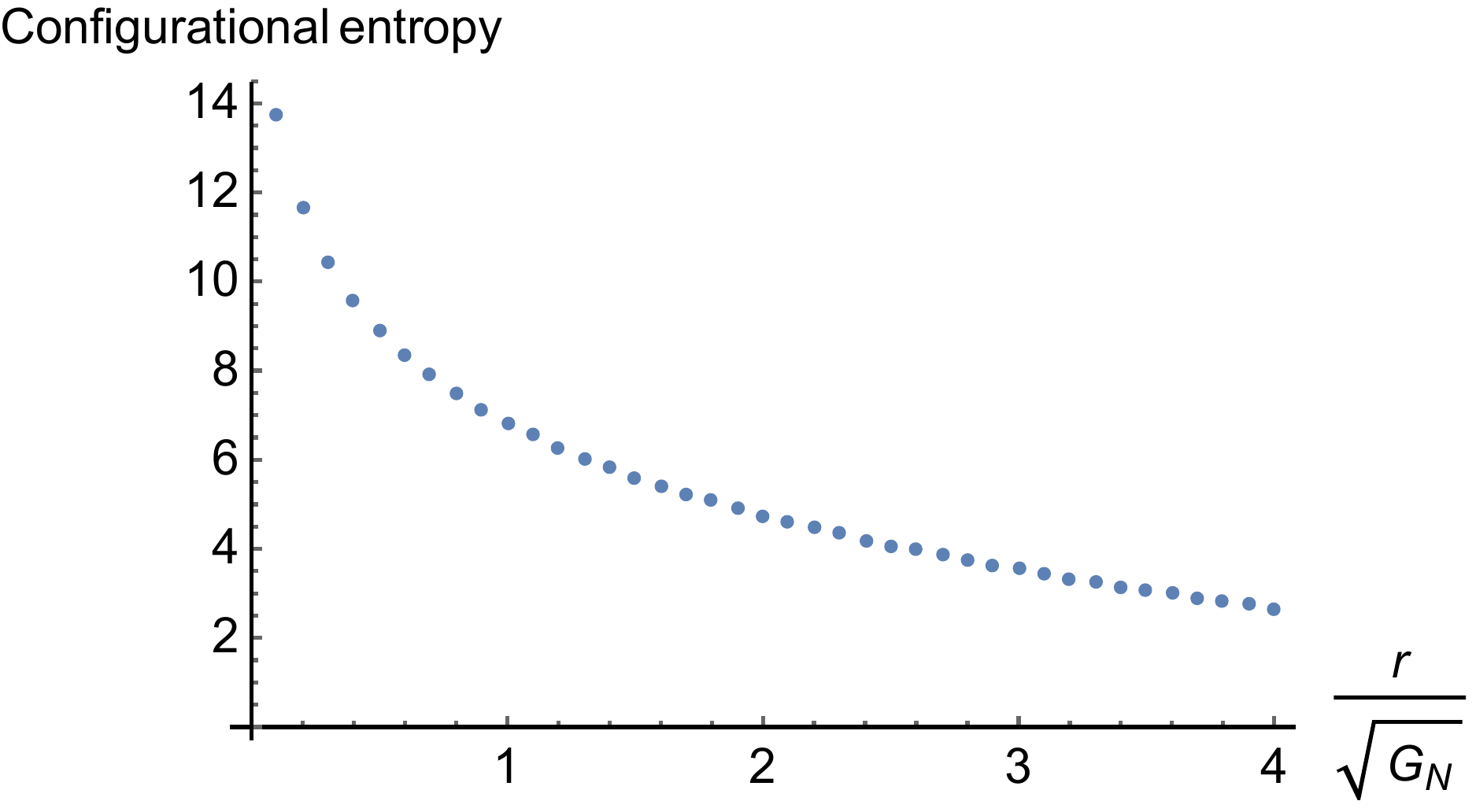}
\caption{Configurational entropy for the flat space black hole as a function of the horizon, in units of $\sqrt{G_N}$.}
\label{f2}
\end{figure}
One can clearly realize that the smaller the black hole, the larger the configurational entropy is.  Smaller black holes are well known to evaporate faster, being more unstable. Hence, our results are fully consistent with the interpretation that the configurational entropy comprises stability of physical systems. States with lesser configurational entropy are more stable. 
  
Similarly to the AdS-Schwarzschild black hole case, the logarithm of the configurational entropy can be studied as a function of the black hole radius, to study whether a scaling relation between the configurational entropy and the flat space black hole radius hold. Our results are depicted in Fig. 4, in what follows.  
 \begin{figure}[H]
\centering\includegraphics[width=7.9cm]{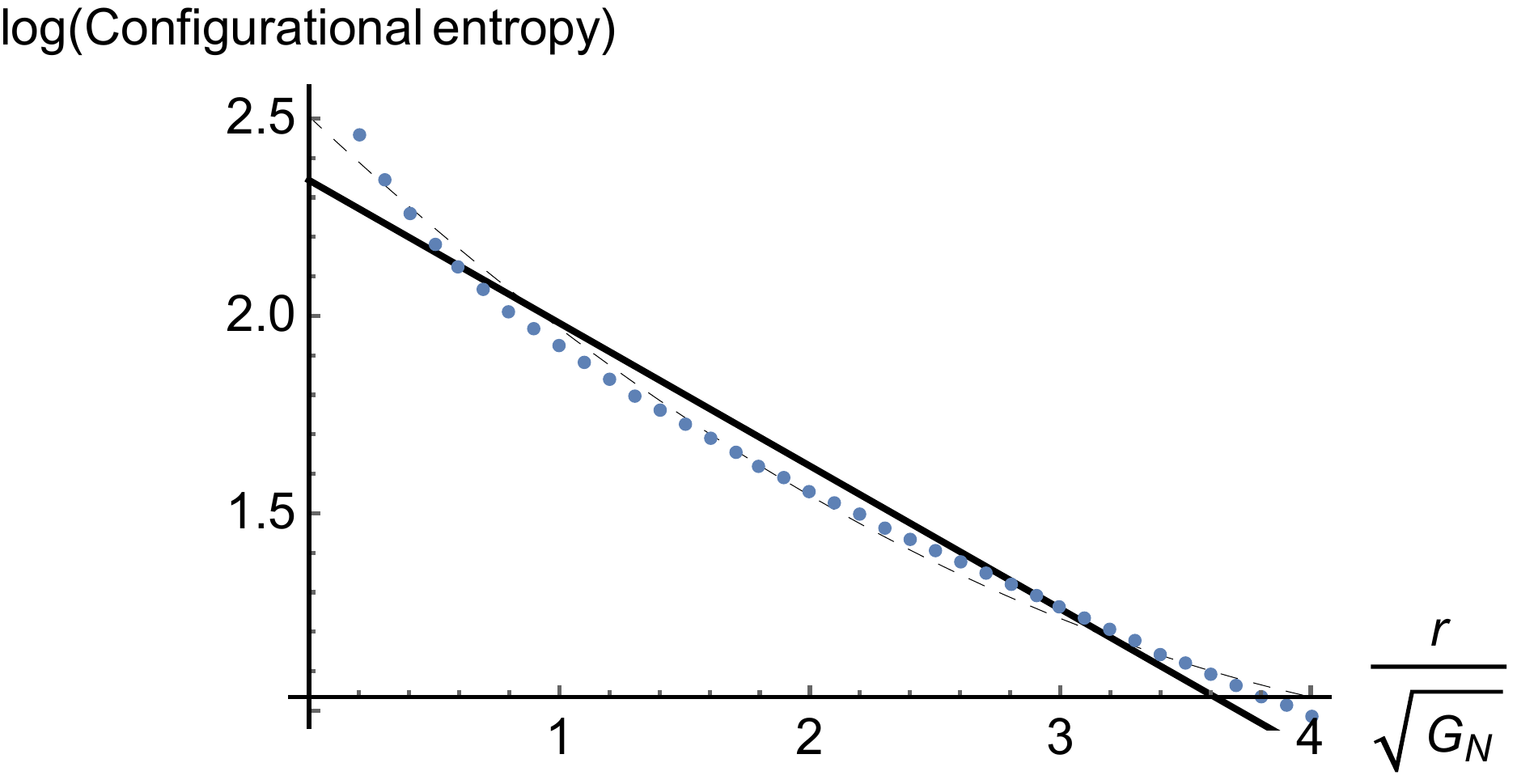}
\caption{Log of the configurational entropy for the flat space black hole as a function of the event horizon in AdS radius units (gray list plot); the black line refers to the linear regression and the dashed line represents a quadratic fit.}
\label{f55}
\end{figure}
The linear regression makes the logarithm of the configurational entropy at the list plot in Fig. 4 to be expressed 
as the monic polynomial  $2.3438 - 0.3620 \frac{r}{\sqrt{G_N}}$. Besides, a quadratic fit to the list plot reads $2.5051  - 0.5925 \frac{r}{\sqrt{G_N}} + 0.0562\frac{r^2}{G_N}$, that is noticed not to be enough to describe the list plot. In particular, the quadratic coefficient is significant and the linear regression 
is not appropriate to derive a scaling relation between the configurational entropy and the flat space black hole event horizon radius.
    
\section{Concluding remarks and outlook} 
 We have shown in this letter that it is possible to find a consistent definition for the configurational entropy of Schwarzschild black holes in AdS and in flat spaces. For the AdS case, the black hole has thermal equilibrium with radiation, and does not evaporate. However it may overcome a  phase transition to  thermal AdS space -- without a black hole -- as shown in Ref. \cite{Hawking:1982dh}.
 The situation of the black hole in flat space is quite different from the physical point of view. It is not subject to a Hawking-Page like phase transition. However, it is not  in thermal equilibrium with  the  radiation, and the flat space black hole evaporates. Moreover, the logarithm of the configurational entropy, as a function of the black hole radius $\frac{r_h}{b}$, 
 reveals an essay of  scaling relation for the AdS-Schwarzschild black hole case (Fig. 2), whereas the flat space case 
 does not present any acceptable scaling relation (Fig. 4).
 
 The remarkable outcome of the present article is that for  two very unlike black holes, 
   the configurational entropy is smaller for the more dominant or the more stable geometries, irrespectively of the background space. 
 It is important to emphasize that in the approach developed here we had to introduce an energy density associated with the contribution of the horizon to the mass of the black hole. As an artifact to find a well defined configurational entropy, this density is defined in the inner region to the black hole. This fact does not mean that we are considering the actual inner region to the event horizon.
 We are describing the black hole as the  region of the space outer to the horizon,  in the same way as in Refs. \cite{Hawking:1982dh}
 and \cite{Witten:1998zw}.  It is necessary to introduce a smeared density  inner to the horizon, since the introduction of a configurational entropy requires that the Fourier transform of the energy density is square integrable. On the other hand,  Parseval theorem asserts that
 \begin{eqnarray}
 \int   \vert \,  {\tilde  \rho} ({\vec \omega}) \,  \vert^2 \,d^3 \omega \sim 
 \int \vert \rho({\vec  r}) \vert^2  d^3 r \,,
\end{eqnarray} 
 so that an energy density sharply located at the horizon leads to a non-square integrable Fourier transform,  that leads to an infinite normalization constant $ {\cal N} $ in Eq. (\ref{norma}). 

The configuration entropy was analysed in this letter for two black hole geometries where the energy densities  behave as  $1/r$ inside the horizon.  It would be interesting to investigate other types of black holes, with different energy density distributions, to check 
whether they present an analogous behaviour. 
 Concluding, the results obtained in this article provide a strong indication that the configurational entropy is indeed related  to the dominance and (or) stability of diverse physical systems.

\medbreak
  
\noindent {\bf Acknowledgments:}    The authors thank to Dr A. Ballon-Bayona 
for important discussions. 
  NRFB is partially supported by CNPq (grant  307641/2015-5). RdR is grateful to CNPq (grant No. 303293/2015-2), to FAPESP (grant 2015/10270-0).

 \end{document}